\documentclass[aps,prd,preprint,superscriptaddress,nofootinbib,amsmath,showpacs]{revtex4-1}
\usepackage{chay,graphicx,bm}
\newcommand{\W}{\mathcal{W}}
\newcommand{\Y}{\mathcal{Y}}
\newcommand{\euv}{\epsilon_{\mathrm{UV}}}
\newcommand{\eir}{\epsilon_{\mathrm{IR}}}

\begin{document}

\title{$\mathcal{N}=4$ Supersymmetric Yang-Mills theory \\
in soft-collinear effective theory}

\def\KU{Department of Physics, Korea University, Seoul 136-701, Korea}

\author{Junegone Chay}
\email[E-mail:]{chay@korea.ac.kr}
\author{Jae Yong Lee}
\email[E-mail:]{littlehiggs@korea.ac.kr}
\affiliation{\KU}

\begin{abstract} \baselineskip 3.0 ex 
We formulate $\mathcal{N}=4$ supersymmetric Yang-Mills theory in terms of soft-collinear effective theory. 
The effective Lagrangian in soft-collinear
effective theory is developed according to the
power counting by a small parameter $\eta \sim p_{\perp}/Q$. 
All the particles in this theory are in the adjoint representation of the $SU(N)$ gauge group, and 
we derive the collinear gauge-invariant Lagrangian in the adjoint and fundamental representations respectively. 
 We consider collinear and ultrasoft Wilson lines in this theory, and show the 
ultrasoft factorization of the collinear Lagrangian by redefining the collinear fields with the use of the ultrasoft Wilson lines.
The vertex correction for a vector fermion current at one loop is explicitly presented as an example 
to illustrate how the computation is  performed in the effective theory.
\end{abstract}

\pacs{11.10.Gh, 11.15.Bt, 11.30.Pb}

\maketitle

\baselineskip 3.0 ex 

\section{Introduction}
The divergence structure of massless or massive gauge theories in the high-energy scattering amplitudes 
has been studied intensively from various perspectives. It has been
considered in quantum chromodynamics (QCD) \cite{Catani:1998bh,Aybat:2006mz}, its effective theory version called soft-collinear 
effective theory (SCET) \cite{Becher:2009qa,Becher:2009cu,Becher:2009kw},  and in AdS/CFT 
approach via $\mathcal{N}=4$ supersymmetric Yang-Mills (SYM) theory \cite{Ferrara:1974pu} for perturbative calculation. 
Each field has its own merit in understanding the divergence structure in high-energy scattering. 
QCD is a robust field since it can be verified by experiment, and the computation is straightforward,
though complicated. The effective theory of QCD for energetic, collinear particles is SCET, and it is formulated
in such a way that the collinear part and the ultrasoft (usoft) part are decoupled from the beginning. Therefore it is more
convenient to track down the sources of ultraviolet (UV) or infrared (IR) divergences in the collinear and the
usoft parts.  The $\mathcal{N}=4$ SYM theory has a lot of symmetries which simplify many
different loop calculations, and especially the theory has conformal invariance \cite{Poggio:1977ma}, that is, the coupling constant does not 
run. Furthermore, duality due to the AdS/CFT correspondence enables to relate some nonperturbative quantities to the corresponding 
quantities in the perturbative region.

The theme of this paper is to see if we can obtain deeper understanding on the divergence structure of high-energy
processes in $\mathcal{N}=4$ SYM theory using the transparent factorization property in SCET. 
In this paper, we construct an effective field theory for 
 $\mathcal{N}=4$ SYM theory in the framework of SCET. The main advantage
of constructing the SCET is the manifest realization of the factorization of collinear and soft contributions. 
The divergence structure at higher 
loops can be categorized into the collinear and the soft divergences, which makes the classification of the 
divergences manifest and enhances
the understanding of the divergence structure of the theory. This paper is the first step toward this goal 
by constructing the effective Lagrangian and studying its properties.

In addition to the manifest factorization property in SCET, the SCET formulation of  
the $\mathcal{N}=4$ SYM theory itself is interesting in various theoretical aspects. First 
the SCET formulation is extended to Weyl fermions. In the original SCET for QCD, a collinear 
Dirac fermion is employed, but it is effectively a 
two-dimensional field once the projection into a collinear sector is performed. The Weyl representation is another
form of the two-dimensional description of fermions, and they may be related, but the physical implications 
are different in different representations. Here we present the SCET Lagrangian in Weyl representation for fermions.
Secondly, all the particles are in the adjoint representation of the $SU(N)$ gauge group. 
This makes the treatment of the group theory factors simple.
In applying SCET to LHC phenomenology, many processes
have been considered involving various particles in different representations \cite{Manohar:2006ga,Beneke:2009rj,Idilbi:2010rs}, 
such as quark-quark scattering,
gluon-gluon scattering, gluon to color-octet scalar particles, etc.. In these cases, different color factors are involved in 
different processes, but we only have to consider the adjoint representation 
in $\mathcal{N}=4$ SYM theory.

On the other hand, the SCET formulation can cast interesting theoretical questions. For example, we can consider
whether the conformal symmetry is realized in an effective theory in which the Lagrangian is reorganized 
order by order in powers of a small parameter. And in the SCET for QCD, a general gauge transformation 
is categorized into collinear, usoft gauge transformations according to how the transformed fields scale and we 
require that a physical quantity should be invariant under both collinear and usoft gauge transformations. 
In $\mathcal{N}=4$ SYM theory, there is another symmetry, that is, supersymmetry. 
It will be interesting if we can also divide the classes of the supersymmetry transformations such that
a physical quantity is invariant under the subgroups of the transformations. This is beyond the scope of
this paper and it will be pursued in the future.

The basic idea of SCET \cite{Bauer:2000ew,Bauer:2000yr,Bauer:2001yt,Chay:2002vy}
starts from the observation that the momentum of an energetic collinear particle 
in the lightcone direction $n^{\mu}$ can be decomposed into
\begin{equation}
p^{\mu} =\overline{n}\cdot p \frac{n^{\mu}}{2} +p_{\perp}^{\mu} +n\cdot p\frac{\overline{n}^{\mu}}{2} 
\sim \mathcal{O}(Q) + \mathcal{O}(Q\eta) +\mathcal{O} (Q\eta^2),
\end{equation}
where $n^{\mu}$, $\overline{n}^{\mu}$ are lightcone vectors satisfying $n^2 =\overline{n}^2 =0$, and 
$n\cdot \overline{n} =2$. The scale $Q$ denotes a large energy characteristic of the 
high-energy scattering, and $\eta = p_{\perp}/\overline{n}\cdot p$ 
is a small parameter, and all the physical observables are expressed in powers of this small parameter $\eta$.
The usoft momentum is given by
\begin{equation}
p_{\mathrm{us}}^{\mu} = (\overline{n}\cdot p_{\mathrm{us}}, p_{\mathrm{us}\perp}^{\mu}, n\cdot 
 p_{\mathrm{us}} )\sim Q(\eta^2, \eta^2, \eta^2).
\end{equation}
Therefore when an usoft particle interacts with a collinear particle, the momentum scaling behavior of a collinear
particle is unchanged.
In QCD, if the particles are on the mass shell, $p^2 \sim Q^2 \eta^2$, and 
$\eta\sim \Lambda_{\mathrm{QCD}}/Q$. But we can allow $p^2 \sim E^2$, 
in an intermediate theory like $\mathrm{SCET}_{\mathrm{I}}$, 
where $E \gg \Lambda_{\mathrm{QCD}}$ is some small energy 
compared to $Q$, and $\eta$ becomes of order $E/Q$. 
On the other hand, there is no scale $\Lambda_{\mathrm{QCD}}$ in $\mathcal{N}=4$ SYM theory,
but it suffices to have a small parameter $\eta=E/Q$.
We describe the interaction of the collinear fields and the usoft fields since we focus on the energetic particles
participating in high-energy scattering according to the power counting method.

The paper is organized as follows: In Sec.~\ref{sym}, we briefly review the Lagrangian in $\mathcal{N}=4$ 
SYM theory. The SCET Lagrangian for fermions in Weyl representation and 
scalars is derived in Sec.~\ref{scetl}. In Sec.~\ref{cowil}, we describe a collinear Wilson line, and its properties. 
In Sec.~\ref{ufac}, we redefine the collinear fields using the usoft Wilson lines to decouple the usoft 
interaction from the collinear fields. The leading collinear Lagrangian after the redefinition is presented, which
explicitly shows this decoupling. In Sec.~\ref{appl}, we consider the vertex correction for a fermion vector current
as an example to show how the collinear and the usoft contributions are computed 
at one loop respectively. We also delineate
the procedure on how to obtain the Wilson coefficients, and the scaling behavior of the operator in SCET.
 
We presume that the readers consist of those who are well versed in $\mathcal{N}=4$ SYM 
theory, but with no knowledge on SCET, or those with the opposite background. The style
of this paper may be easily readable for the latter, and we try to fill the gap as much as possible to make the
paper understandable for the former. The detailed SCET calculations will appear in Appendix, not to 
interfere with the logical flow of the paper. 

\section{$\mathcal{N}=4$ supersymmetric Yang-Mills Lagrangian\label{sym}}
The Lagrangian for $\mathcal{N}=4$ SYM theory with the $SU(N)$ gauge group is given by \cite{Henn:2009bd}
\begin{eqnarray} \label{susyl}
\mathcal{L} &=&\mathrm{Tr} \Bigl( -\frac{1}{4} G_{\mu\nu}G^{\mu\nu} 
+ \overline{\lambda}_i \overline{\sigma}^{\mu} iD_{\mu} \lambda_i
+\frac{1}{2} D_{\mu} \phi_{ij} D^{\mu} \phi^{ij} \nonumber \\
&&-ig\lambda_i  [\lambda_j,\phi^{ij}] -ig \overline{\lambda}^i [\overline{\lambda}^j,\phi_{ij}] 
+\frac{g^2}{4} [\phi_{ij}, \phi_{kl}][\phi^{ij}, \phi^{kl}]\Bigr), 
\end{eqnarray}
where $\phi^{ij} =-\phi^{ji}$ and the indices $i$, $j$, $k$, $l$ run from 1 to 4. Due to supersymmetry, the only 
coupling constant that appears in the Lagrangian is the gauge coupling $g$. The field contents of the theory can 
be classified in terms of the supersymmetric properties, but here it suffices to specify the fields according to 
the properties in Lorentz transformation.
Here $G_{\mu\nu}$ are the field strength tensor for the $SU(N)$ gauge fields, 
$\lambda_i$ are the adjoint Weyl fermion fields, 
and $\phi_{ij}$ are the scalar fields. 

All the fields in the Lagrangian are in the $SU(N)$ adjoint representations.
From now on, we will drop all the particle flavor indices $i$ and $j$ since they are irrelevant in the SCET formulation, but can
be inserted at the end in a straightforward way. In Eq.~(\ref{susyl}), the Lagrangian contains trace, which means that we write
$\lambda = \lambda^a t^a$, $\phi=\phi^a t^a$,  and $A^{\mu} = A^{\mu a} t^a$, 
where $t^a$ is the $SU(N)$ generators in the fundamental representation 
(i. e. $N\times N$ matrix).  In this case, the covariant derivative applied to a fermion is defined as
\begin{equation}
D^{\mu} \lambda = \partial^{\mu} \lambda  -ig [A^{\mu},\lambda].
\end{equation}
In terms of the color components, or in the adjoint representation, the interaction of $\lambda$ with a gluon is given as
\begin{eqnarray}
\mathcal{L}_{\mathrm{int}} &=& \mathrm{Tr}\, \Bigl(g \overline{\lambda} \overline{\sigma}^{\mu} A_{\mu} 
\lambda \Bigr) 
= g \overline{\lambda}_a \overline{\sigma}^{\mu} A_{\mu b} \lambda_c \mathrm{Tr}\, 
\Bigl( t^a [t^b,t^c]\Bigr)   \nonumber \\
&=&igf^{abc} T_F \overline{\lambda}_a \overline{\sigma}\cdot A_b \lambda_c =\frac{g}{2} 
\overline{\lambda} \overline{\sigma}\cdot \mathcal{A}  \lambda,
\end{eqnarray}
where $T_F=1/2$ for $SU(N)$, and $\mathcal{A}^{\mu} = A^{\mu b} T^b$ with the 
adjoint representation $(T^b)_{ac} =-if^{bac}$. The Lagrangian can be written in either way, and 
the expression in Eq.~(\ref{susyl}) is the conventional one in the study of $\mathcal{N}=4$ SYM 
theory. However, the relations between the two expressions will be explored in detail in this paper.
We will follow the conventions of Ref.~\cite{Dreiner:2008tw} for the metric and the representations of fermions.

\section{SCET Lagrangian\label{scetl}}
\subsection{Collinear fermion Lagrangian in Weyl representation}

First we consider the Lagrangian for the Weyl fermion $\lambda$, and we need to construct the SCET 
for the two-dimensional Weyl fields. In SCET for QCD, a collinear Dirac fermion $\xi_n$ is employed, 
but they are actually described 
by the two-dimensional spinors using the projection operators. The Dirac fermion field $\psi$ in QCD is expressed
in terms of the $n$-collinear field $\xi_n$ and the $\overline{n}$-collinear field $\xi_{\bar{n}}$ as
\begin{equation}
\psi (x) = \sum_{\tilde{p}} e^{-i\tilde{p}\cdot x} \Bigl( \xi_n (x) +\xi_{\bar{n}}(x)\Bigr), 
\end{equation}
where $\tilde{p}^{\mu} = \overline{n}\cdot p n^{\mu}/2 + p_{\perp}^{\mu}$ is the label momentum, of 
the order of $Q$ and $Q\eta$. Once the 
label momentum is extracted, the resulting Lagrangian describes the dynamics with the fluctuation of order $Q\eta^2$.
The effective fields $\xi_n$ and $\xi_{\bar{n}}$ satisfy the relations 
$\FMslash{n} \xi_n =0$, $\FMslash{\overline{n}} \xi_{\bar{n}}=0$ and 
\begin{eqnarray}
&&\frac{\FMslash{n} \FMslash{\overline{n}}}{4} \xi_n =\xi_n,  
\ \frac{\FMslash{\overline{n}}\FMslash{n}}{4} \xi_n =0, \nonumber \\
&& \frac{\FMslash{n} \FMslash{\overline{n}}}{4} \xi_{\bar{n}} =0,  
\ \frac{\FMslash{\overline{n}}\FMslash{n}}{4} \xi_{\bar{n}} =\xi_{\bar{n}}. 
\end{eqnarray}
Here $P_n = \FMslash{n}\FMslash{\overline{n}}/4$ and 
$P_{\bar{n}} =\FMslash{\overline{n}} \FMslash{n}/4$ act as projection operators satisfying 
$P_n^2 =P_n$, $P_{\bar{n}}^2 =P_{\bar{n}}$, $P_n P_{\bar{n}}= P_{\bar{n}} P_n=0$ 
and $P_n+P_{\bar{n}}=1$.
Therefore the Dirac fermions in SCET are effectively described by two-dimensional spinors. But 
here we express the Lagrangian in another two-dimensional spinor
representation, that is, the Weyl representation. We will focus on the left-handed 
Weyl fields, and the case with the right-handed fields can be 
extended in a straightforward way. 

The starting point is to use the gamma matrices in Weyl representation, which are given by
\begin{equation}
\gamma^{\mu} =\begin{pmatrix} 0&\sigma^{\mu} \cr
                                                    \overline{\sigma}^{\mu}&0
                           \end{pmatrix},
\end{equation}
where  $\sigma^{\mu} = (1,\bm{\sigma})$,  and 
$\overline{\sigma}^{\mu} = (1,-\bm{\sigma})$ with the Pauli matrices $\bm{\sigma}$. 
A Dirac fermion $\psi$ can be written in terms of the left-handed and the right-handed 
Weyl fields $\psi_L$ and $\psi_R$ as
\begin{equation}
\psi =\begin{pmatrix} \psi_L \cr \psi_R\end{pmatrix}.
\end{equation}
The fields $\psi_L$ and $\psi_R$ transform under an infinitesimal Lorentz transformation as
\begin{eqnarray}
\psi_L &\rightarrow& \Bigl( 1 -i \bm{\theta} \cdot \frac{\bm{\sigma}}{2} 
-\bm{\beta}\cdot \frac{\bm{\sigma}}{2} \Bigr) \psi_L, \nonumber \\
 \psi_R &\rightarrow& \Bigl( 1 -i \bm{\theta} 
\cdot\frac{\bm{\sigma}}{2} +\bm{\beta} \cdot 
\frac{\bm{\sigma}}{2}\Bigr) \psi_R,
\end{eqnarray}
where $\bm{\theta}$ ($\bm{\beta}$) 
denotes the infinitesimal rotation (boost). 
The free fields satisfy the equation of motion $i\overline{\sigma}\cdot \partial \psi_L=0$, 
$i \sigma \cdot \partial \psi_R=0$.

For an energetic particle moving in the $n$ direction, the momentum scales as
\begin{equation}
p^{\mu} = (\overline{n}\cdot p, p_{\perp}, n\cdot p)\sim Q(1,\eta,\eta^2),
\end{equation}
where $\eta = p_{\perp}/\overline{n}\cdot p$ is a small parameter. The fermion $\lambda$ in 
the full theory can be written as 
\begin{equation} \label{decom}
\lambda (x) = \sum_{\tilde{p}} e^{-i\tilde{p}\cdot x}\lambda_q (x),
\end{equation}
where the label momentum $\tilde{p}^{\mu} = \overline{n}\cdot p n^{\mu}/2 +p_{\perp}^{\mu}$ is extracted. 
The field $\lambda_q$ can be
decomposed into $\lambda_q = \lambda_n +\lambda_{\bar{n}}$, where $\lambda_n$ and 
$\lambda_{\bar{n}}$ are given by
\begin{equation}
\lambda_n = \frac{1}{4}n\cdot \sigma \overline{n}\cdot \overline{\sigma} \lambda_q, 
\ \lambda_{\bar{n}} = \frac{1}{4} 
\overline{n}\cdot \sigma n\cdot\overline{\sigma} \lambda_q.
\end{equation}
These fields satisfy the relation $n\cdot \overline{\sigma} \lambda_n =0$, $\overline{n}\cdot \overline{\sigma}
\lambda_{\bar{n}}=0$.

The matrices $P_n^L =n\cdot \sigma \overline{n} \cdot \overline{\sigma}/4$ and 
$P_{\bar{n}}^L=\overline{n} \cdot \sigma n\cdot  \overline{\sigma}/4$ act as 
projection operators into the $n$ and $\overline{n}$ left-handed collinear fermions in Weyl representation. 
This can be verified by representing the usual projection operators
for Dirac fermions $\FMslash{n} \FMslash{\overline{n}}/4$ and $\FMslash{\overline{n}}\FMslash{n}/4$ 
in Weyl representation as
\begin{equation}
\frac{\FMslash{n}\FMslash{\overline{n}}}{4} 
=\frac{1}{4}\begin{pmatrix} n\cdot \sigma \overline{n} \cdot \overline{\sigma} & 0\cr
   0& n\cdot \overline{\sigma} \overline{n}\cdot \sigma \end{pmatrix}, \ 
\frac{\FMslash{\overline{n}}\FMslash{n}}{4} = \frac{1}{4}
\begin{pmatrix} \overline{n}\cdot \sigma n \cdot \overline{\sigma} & 0\cr
    0& \overline{n}\cdot \overline{\sigma} n\cdot \sigma \end{pmatrix}.
\end{equation}
The diagonal matrices correspond to the projection operators for left-handed and right-handed 
fields respectively, which are given as
\begin{eqnarray}
&&P_n^L =\frac{1}{4} n\cdot \sigma \overline{n} \cdot \overline{\sigma}, 
\  P_{\bar{n}}^L=\frac{1}{4} \overline{n} \cdot \sigma n\cdot  \overline{\sigma} \nonumber \\
&&P_n^R= \frac{1}{4} n\cdot \overline{\sigma} \overline{n} \cdot \sigma, 
\  P_{\bar{n}}^R=\frac{1}{4} \overline{n} \cdot \overline{\sigma} n\cdot  \sigma.
\end{eqnarray}
The projection operators satisfy the properties $(P_n^{L,R})^2 =P_n^{L,R}$, 
$(P_{\bar{n}}^{L,R})^2 =P_{\bar{n}}^{L,R}$, $P_n^{L,R} P_{\bar{n}}^{L,R}=0$, and $P_n^{L,R} +P_{\bar{n}}^{L,R}=1$. 
This can be verified explicitly using the identity
\begin{equation}
\sigma^{\mu}\overline{\sigma}^{\nu} +\sigma^{\nu} \overline{\sigma}^{\mu} = 2 g^{\mu\nu}, \ 
\overline{\sigma}^{\mu} \sigma^{\nu} + \overline{\sigma}^{\nu} \sigma^{\mu} = 2g^{\mu\nu}.
\end{equation}

Using the decomposition in Eq.~(\ref{decom}), the Lagrangian for the left-handed fermions 
 can be written as
\begin{eqnarray}
\mathcal{L}_{\lambda} &=& \mathrm{Tr}\, \Biggl(
\overline{\lambda}_n \frac{\overline{n}\cdot \overline{\sigma}}{2} n\cdot iD\lambda_n + \overline{\lambda}_n 
\overline{\sigma}\cdot (p_{\perp}+iD_{\perp}) \lambda_{\bar{n}} \nonumber \\
&+&\overline{\lambda}_{\bar{n}}\overline{\sigma} \cdot (p_{\perp}
+iD_{\perp}) \lambda_n +\overline{\lambda}_{\bar{n}} \frac{n\cdot \overline{\sigma}}{2} 
(\overline{n}\cdot p +\overline{n}\cdot iD)\lambda_{\bar{n}} \Biggr).
\end{eqnarray}
The equation of motion $\partial \mathcal{L}/\partial \overline{\lambda}_{\bar{n}} =0$ reads
\begin{equation}
\overline{\sigma}\cdot  (p_{\perp}+iD_{\perp}) \lambda_n +\frac{n\cdot \overline{\sigma}}{2} 
(\overline{n}\cdot p +\overline{n}\cdot iD) \lambda_{\bar{n}} =0,
\end{equation}
from which $\lambda_{\bar{n}}$ is given by
\begin{equation}
\lambda_{\bar{n}}= -\frac{\overline{n}\cdot \sigma}{2} \frac{1}{\overline{n}\cdot p 
+\overline{n}\cdot iD} \overline{\sigma} \cdot (p_{\perp}+iD_{\perp}) \lambda_n.
\end{equation}
However, it is not clear how to apply the covariant derivative in the denominator to the fermion. In order to see how it
works, we write the Lagrangian in the adjoint representation as
\begin{eqnarray}
\mathcal{L}_{\lambda} &=& \frac{1}{2} \Biggl[ \overline{\lambda}_n^a \frac{\overline{n}\cdot \overline{\sigma}}{2}
n\cdot i\partial \lambda_n^a +gf^{abc} \overline{\lambda}_n^a \frac{\overline{n}\cdot \overline{\sigma}}{2}
n\cdot A^b \lambda_n^c
+ \overline{\lambda}_n^a \overline{\sigma} \cdot (p_{\perp} + i\partial_{\perp}) \lambda_n^a
+gf^{abc}\overline{\lambda}_n^a \overline{\sigma}\cdot A_{\perp}^b \lambda_n^c   \nonumber \\
&+& \overline{\lambda}_n^a \overline{\sigma} \cdot (p_{\perp} +i\partial_{\perp}) \lambda_{\bar{n}}^a 
+gf^{abc} \overline{\lambda}_n^a \overline{\sigma}\cdot A_{\perp}^b \lambda_{\bar{n}}^c
+\overline{\lambda}_{\bar{n}}^a \overline{\sigma} \cdot (p_{\perp} +i\partial_{\perp}) \lambda_n^a 
+gf^{abc} \overline{\lambda}_{\bar{n}}^a \overline{\sigma}\cdot A_{\perp}^b \lambda_n^c \nonumber \\
&+&\overline{\lambda}_{\bar{n}}^a \frac{n\cdot \overline{\sigma}}{2} (\overline{n}\cdot p 
+\overline{n}\cdot i\partial)
\lambda_{\bar{n}}^a+gf^{abc} \overline{\lambda}_{\bar{n}}^a   \frac{n\cdot \overline{\sigma}}{2}
\overline{n}\cdot A^b \lambda_{\bar{n}}^c\Biggr]
\nonumber \\
&=& \frac{1}{2}\Biggl[ \overline{\lambda}_n \frac{\overline{n}\cdot \overline{\sigma}}{2} in\cdot 
\mathcal{D} \lambda_n+\overline{\lambda}_n \overline{\sigma}\cdot (p_{\perp} +i\mathcal{D}_{\perp}) 
\lambda_{\bar{n}} \nonumber \\ 
&&+\overline{\lambda}_{\bar{n}} \overline{\sigma}\cdot (p_{\perp} +i\mathcal{D}_{\perp}) 
\lambda_n +\overline{\lambda}_{\bar{n}}\frac{n\cdot \overline{\sigma}}{2}
 (\overline{n}\cdot p +\overline{n}\cdot i\mathcal{D})
\lambda_{\bar{n}} \Biggr],
\end{eqnarray}
where $\mathcal{D}^{\mu} =\partial^{\mu} -igA^{\mu a}(T^a)$, and $(T^a)_{bc} =-if^{abc}$ is the adjoint
representation. Now we can use the equation of motion 
$\partial \mathcal{L}_{\lambda} /\partial \overline{\lambda}_{\bar{n}}^a=0$,  which yields
\begin{equation}
\overline{\sigma}\cdot (p_{\perp} +i\mathcal{D}_{\perp}) \lambda_n +\frac{n\cdot \overline{\sigma}}{2} 
(\overline{n}\cdot p +\overline{n}\cdot i\mathcal{D}) \lambda_{\bar{n}} =0,
\end{equation}
and $\lambda_{\bar{n}}$ is given by
\begin{equation} \label{subad}
\lambda_{\bar{n}}= -\frac{\overline{n}\cdot \sigma}{2} \frac{1}{\overline{n}\cdot p 
+\overline{n}\cdot i\mathcal{D}} \overline{\sigma} \cdot (p_{\perp}+i\mathcal{D}_{\perp}) \lambda_n,
\end{equation}
where there is no ambiguity in applying the covariant derivative operator in the denominator.
From this expression, we can see that $\lambda_{\bar{n}}$ is a small component suppressed by 
$\eta$ compared to $\lambda_n$.

Let us decompose the gauge field into the collinear and the usoft gauge fields as $\mathcal{A}^{\mu} 
= \mathcal{A}_n^{\mu} +\mathcal{A}_{\mathrm{us}}^{\mu}$, where the gauge fields scale as
\begin{eqnarray} \label{gauge}
\mathcal{A}_n^{\mu} &=& (\overline{n} \cdot \mathcal{A}_n, \mathcal{A}_{n,\perp}^{\mu}, n\cdot
\mathcal{A}) \sim Q(1, \eta, \eta^2), \nonumber \\
\mathcal{A}_{\mathrm{us}}^{\mu} &=& (\overline{n} \cdot \mathcal{A}_{\mathrm{us}}, 
\mathcal{A}_{\mathrm{us},\perp}^{\mu}, n\cdot \mathcal{A}_{\mathrm{us}}) \sim Q(\eta^2, \eta^2, \eta^2).
\end{eqnarray}
Only the collinear and the usoft gauge fields can interact with collinear fermions, otherwise the momentum scaling behavior
is violated.
The collinear and usoft gauge fields are the subsets of the original gauge fields in the full theory, with the definite scaling
behavior. And the gauge transformations in the full theory can also be divided into the collinear and usoft gauge 
transformations, under which the scaling behavior of each field in SCET is preserved.

\begin{figure}[t]
\begin{center} 
\includegraphics[height=5.5cm]{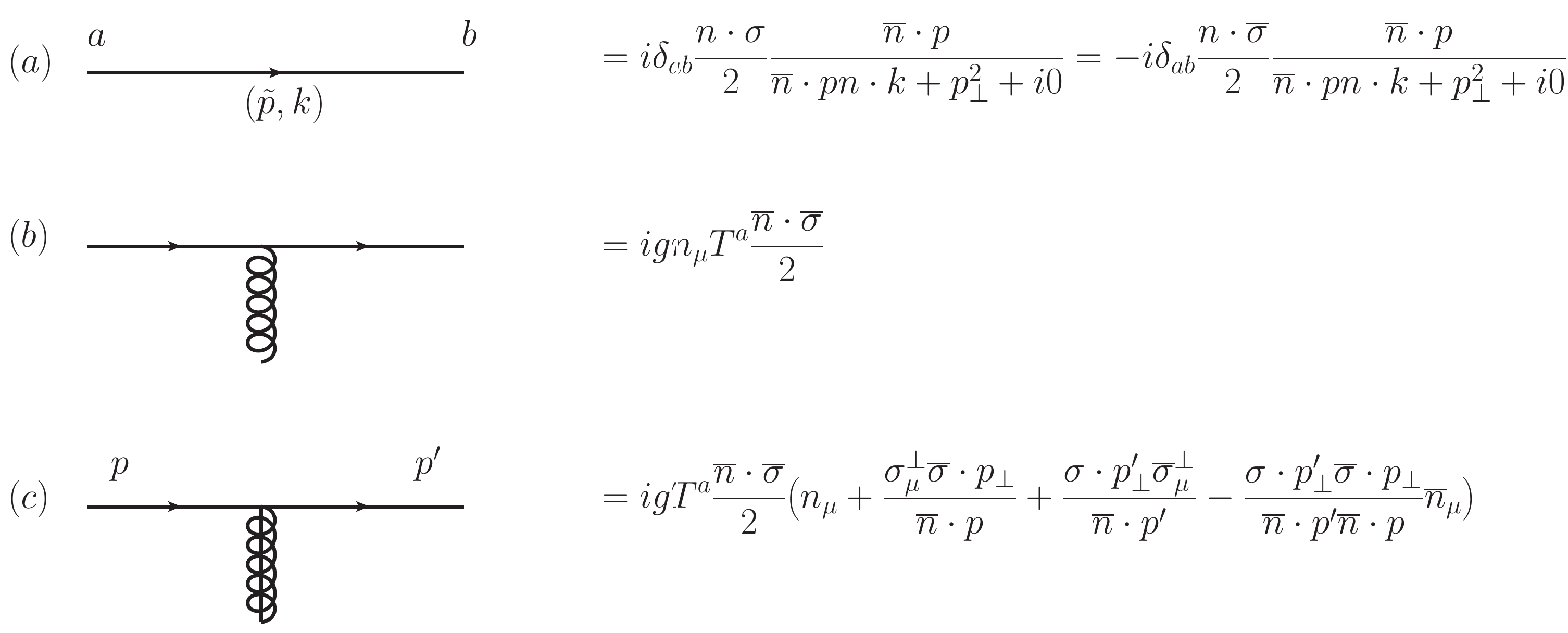}
\end{center}\vspace{-.5cm}
\caption{\baselineskip 3.0ex 
Feynman rules for $\mathcal{L}_{\lambda}^{(0)}$ to order $g$ in SCET. (a) collinear fermion propagator with label 
momentum $\tilde{p}$ and residual momentum $k$, (b) collinear fermion interaction with a usoft gauge field, and (c)
collinear fermion interaction with a collinear gauge field. Here $(T^a)_{bc} =-if^{abc}$ is the adjoint $SU(N)$ 
generators.}
\label{feynman}
\end{figure}

Integrating out $\lambda_{\bar{n}}$ using the equation of motion, the Lagrangian for the left-handed fermion $\lambda_n$ 
in the $n$ direction at leading order  in $\eta$ is written as
\begin{eqnarray} \label{lzero}
\mathcal{L}_{\lambda}^{(0)} &=& \frac{1}{2} \overline{\lambda}_n \frac{\overline{n}\cdot \overline{\sigma}}{2}
\Bigl(n\cdot i\mathcal{D}+ \sigma\cdot \mathcal{D}_{c\perp}
\frac{1}{\overline{n}\cdot \mathcal{P} +g\overline{n}\cdot \mathcal{A}_n} \overline{\sigma} \cdot 
\mathcal{D}_{c\perp} \Bigr) \lambda_n,
\end{eqnarray}
where $n\cdot \mathcal{D} = n\cdot \partial -ig(n\cdot \mathcal{A}_n +n\cdot \mathcal{A}_{\mathrm{us}})$ 
includes the usoft gauge field because $n\cdot \mathcal{A}_{\mathrm{us}}$ has the same power counting $Q\eta^2$
as  $n\cdot  \mathcal{A}_n$, and 
$\mathcal{D}_{c\perp} ^{\mu}= \mathcal{P}_{\perp}^{\mu} +g\mathcal{A}_{n\perp}^{\mu}$ is the collinear
covariant derivative. The operators $\overline{n}\cdot \mathcal{P}$ and $\mathcal{P}_{\perp}^{\mu}$ extract 
the label momenta $\overline{n}\cdot p$ and $p_{\perp}^{\mu}$ respectively.
 The Lagrangian at higher orders in $\eta$ can be obtained by a systematic Taylor series expansion
in powers of $\eta$, which can involve the usoft covariant derivatives.

The Feynman rules for the Lagrangian $\mathcal{L}_{\lambda}^{(0)}$ to order $g$ are shown in Fig.~\ref{feynman}.
The overall factor $T_F=1/2$ in front of the Lagrangian is neglected here.
The two different expressions for the propagator of a collinear 
fermion in Fig.~\ref{feynman} (a) result from the property of $\sigma$ 
and $\overline{\sigma}$ \cite{Dreiner:2008tw}. The Lagrangian will be cast in a simpler form after 
the collinear Wilson line is introduced in the  next section. Note that the Lagrangian in Eq.~(\ref{lzero}) is written
in the adjoint representation, and the Lagrangian in the fundamental representation will also be presented later.

\subsection{Scalar Lagrangian}
The scalar Lagrangian in full theory is given by
\begin{equation}
\mathcal{L}_{\phi} =\mathrm{Tr}\, \Bigl(\frac{1}{2} D_{\mu} \phi D^{\mu} \phi\Bigr) 
= -\mathrm{Tr}\, \Bigl(\frac{1}{2}  \phi D_{\mu} D^{\mu} \phi\Bigr).
\end{equation}
For an energetic, collinear scalar particle, we define the collinear scalar field by extracting the label momentum 
with the normalization
\begin{equation}
\phi (x) = \sum_{\tilde{p}} \frac{1}{\sqrt{\overline{n}\cdot p}} \Bigl(e^{-i\tilde{p}\cdot x} \phi_n (x)
+e^{i\tilde{p}\cdot x} \phi_n^* (x) \Bigr). 
\end{equation}
Then at leading order in $\eta$, the SCET Lagrangian for the scalar is given by 
\begin{equation} \label{sscet}
\mathcal{L}_{\phi} = \mathrm{Tr}\, \Bigl[\phi_n^* \Bigl( n\cdot iD +\frac{p_{\perp}^2}{\overline{n}\cdot p} \Bigr) 
\phi_n \Bigr]\rightarrow \mathrm{Tr}\, \Bigl[\phi_n^* \Bigl( n\cdot iD +  
\frac{\mathcal{P}_{\perp}^2}{\overline{n}\cdot \mathcal{P}} \Bigr)  \phi_n \Bigr].
\end{equation}

Note that there also exists fermion-scalar interaction in the full Lagrangian of Eq.~(\ref{susyl}). However,
there is no interaction of collinear fermions in SCET with either a collinear scalar or an usoft scalar particle 
at leading order in $\eta$. 
It is due to the structure of the interaction
of massless collinear fermions with scalar particles. In order to see this, let us neglect the flavor and the color 
indices for a moment, and this interaction is of the form $\lambda \lambda \phi$. 
We first project the second $\lambda$ into a collinear fermion
as $\lambda \lambda_n$. Using the projection operator explicitly, it is written as
\begin{eqnarray} \label{nofs}
\lambda \lambda_n &=& \lambda P_n^L \lambda =\lambda \frac{1}{4} n\cdot \sigma 
\overline{n}\cdot \overline{\sigma} \lambda_n
=\frac{1}{4} n_{\mu} \overline{n}_{\nu} \lambda \sigma^{\mu} \overline{\sigma}^{\nu} \lambda_n 
= \frac{1}{4} n_{\mu} \overline{n}_{\nu} \lambda_n \sigma^{\nu} \overline{\sigma}^{\mu} \lambda \nonumber \\
&=&\lambda_n \frac{1}{4} \overline{n}\cdot \sigma n\cdot \overline{\sigma} \lambda 
= \lambda_n P_{\bar{n}}^L \lambda=\lambda_n \lambda_{\bar{n}}. 
\end{eqnarray} 
Here we use the relation 
\begin{equation}
\psi \sigma^{\mu} \overline{\sigma}^{\nu} \chi = \chi \sigma^{\nu} \overline{\sigma}^{\mu} \psi,
\end{equation}
for arbitrary fermion fields $\chi$ and $\psi$. Therefore the scalar-fermion interaction vanishes at leading order and it
begins at subleading order $\eta$  with the presence of a small component $\lambda_{\bar{n}}$. 
The explicit subleading 
scalar-fermion interaction is presented after the collinear Wilson line is introduced in the next section.
This relation holds irrespective of whether the scalar is collinear or usoft.  It is due to the fact that massless fermions conserve chirality. 
This fact greatly simplifies the structure of $\mathcal{N}=4$ SYM theory in SCET as we shall see below.


Finally, the collinear scalar interaction in SCET is obtained by 
replacing $\phi^{ij}$ by the corresponding collinear field $\phi_n^{ij}$. 
Since the collinear Lagrangian for the gauge field is already studied in the literature on SCET \cite{Bauer:2001yt},
we will not derive it here. The usoft Lagrangian is obtained if we replace the full-theory fields by the usoft fields in
the full-theory Lagrangian.
\section{Collinear Wilson line\label{cowil}}
The Lagrangian in Eq.~(\ref{lzero}) can be expressed in a form showing manifest collinear gauge invariance 
by introducing the collinear Wilson line
\begin{equation}\label{colwil}
\W_n =\sum_{\mathrm{perm.}} \exp \Bigl[-g \frac{1}{\overline{n}\cdot \mathcal{P}} \overline{n}
\cdot \mathcal{A}_n\Bigr],
\end{equation}
where the
bracket implies that the operator $\overline{n}\cdot\mathcal{P}$ is applied only inside the bracket.
And $\mathcal{A}_n^{\mu} = A_n^{\mu,a}T_a$ is the $n$-collinear gauge field in the adjoint representation.

The gauge transformation in the full theory can be decomposed into the collinear gauge 
transformation $\mathcal{U}_c$, and the usoft gauge transformations $\mathcal{U}_{\mathrm{us}}$. A collinear
gauge transformation $\mathcal{U}_c (x)= \exp[i\alpha^a (x) T^a]$ is defined as the subset of gauge 
transformations where $\partial^{\mu} \mathcal{U}_c \sim Q( 1,\eta,\eta^2)$. For a collinear gauge transformation
$\mathcal{U}_c (x)$, we extract the large label momentum as was done for collinear fields,
\begin{equation}
\mathcal{U}(x) = \sum_P e^{-iP\cdot x} \mathcal{U}_{Pc} (x),
\end{equation}
where $\partial^{\mu}\mathcal{U}_{Pc} \sim Q\eta^2$. We will drop the label momenta with the understanding that
label momentum indices are arranged to conserve the label momenta.
Usoft gauge transformations
$\mathcal{U}_{\mathrm{us}} (x) =\exp [i\beta_{\mathrm{us}}^a (x) T^a]$ are the subset where $\partial^{\mu}
\mathcal{U}_{\mathrm{us}} (x) \sim Q(\eta^2,\eta^2, \eta^2)$. The characteristics of each gauge transformation for the
SCET for QCD is explained in Ref.~\cite{Bauer:2001yt}. The gauge transformations for the collinear, usoft fields
and the Wilson lines are listed in Table~\ref{gtran}.  The covariant derivatives appearing in the transformation of the 
collinear gauge field is
\begin{equation}
i\tilde{D}^{\mu} \equiv \frac{n^{\mu}}{2} \overline{n}\cdot \mathcal{P} +\mathcal{P}_{\perp}^{\mu}
+\frac{\overline{n}^{\mu}}{2} in\cdot D,
\end{equation}
with $iD^{\mu} = i\partial^{\mu} +gA_{\mathrm{us}}^{\mu}$ where only the usoft field appears.

\begin{table}[t]
\caption{\baselineskip 3.0ex \label{gtran} Gauge transformations for the collinear, usoft fields and 
the Wilson lines. The label momenta are suppressed,
which can be inserted with the label momentum conservation. For each field, the first (second) row corresponds to the 
fundamental (adjoint) representation.}
\begin{ruledtabular}
\begin{tabular}{ccc}
Fields  & Collinear transformation & Usoft transformation \\ \hline
$\lambda_n$ & $U_c \lambda_n U_c^{\dagger}$ & $U_{\mathrm{us}} \lambda_n U_{\mathrm{us}}^{\dagger}$ \\
& $\mathcal{U}_c \lambda_n$ & $\mathcal{U}_{\mathrm{us}} \lambda_n$\\ 
$ A_n^{\mu}$ & $U_c A_n^{\mu} U_c^{\dagger} + \displaystyle\frac{1}{g} U_c [i\tilde{D}^{\mu} U_c^{\dagger}]$& 
$U_{\mathrm{us}}A_n^{\mu} U_{\mathrm{us}}^{\dagger}$\\
$\mathcal{A}_n^{\mu}$ & $\mathcal{U}_c \mathcal{A}_n^{\mu} \mathcal{U}_c +\displaystyle \frac{1}{g}
\mathcal{U}_c [i\tilde{\mathcal{D}}^{\mu} \mathcal{U}_c^{\dagger}]$& $\mathcal{U}_{\mathrm{us}}
\mathcal{A}_n^{\mu} \mathcal{U}_{\mathrm{us}}^{\dagger}$ \\
$\lambda_{\mathrm{us}}$ & $\lambda_{\mathrm{us}}$ & $U_{\mathrm{us}} \lambda_{\mathrm{us}} 
U_{\mathrm{us}}^{\dagger}$ \\
& $\lambda_{\mathrm{us}}$ & $\mathcal{U}_{\mathrm{us}} \lambda_{\mathrm{us}}$ \\
$A_{\mathrm{us}}^{\mu}$& $A_{\mathrm{us}}^{\mu}$ & $U_{\mathrm{us}} \Bigl( A_{\mathrm{us}}^{\mu}
+\displaystyle \frac{i}{g} \partial^{\mu} \Bigr) U_{\mathrm{us}}^{\dagger}$ \\
$\mathcal{A}_{\mathrm{us}}^{\mu}$& $\mathcal{A}_{\mathrm{us}}^{\mu}$ &
$\mathcal{U}_{\mathrm{us}} \Bigl( \mathcal{A}_{\mathrm{us}}^{\mu}
+\displaystyle \frac{i}{g} \partial^{\mu} \Bigr) \mathcal{U}_{\mathrm{us}}^{\dagger}$ \\
Wilson lines && \\ \hline
$W$ & $U_c W$& $U_{\mathrm{us}} WU_{\mathrm{us}}^{\dagger}$ \\
$\W$ & $\mathcal{U}_c \W$& $\mathcal{U}_{\mathrm{us}}\W \mathcal{U}_{\mathrm{us}}^{\dagger}$\\
$Y$ & $Y$ &$U_{\mathrm{us}}Y$ \\
$\Y$ & $\Y$ & $\mathcal{U}_{\mathrm{us}} \Y$\\
\end{tabular}
\end{ruledtabular}
\end{table}

\begin{figure}[t]
\begin{center} 
\includegraphics[height=3.cm]{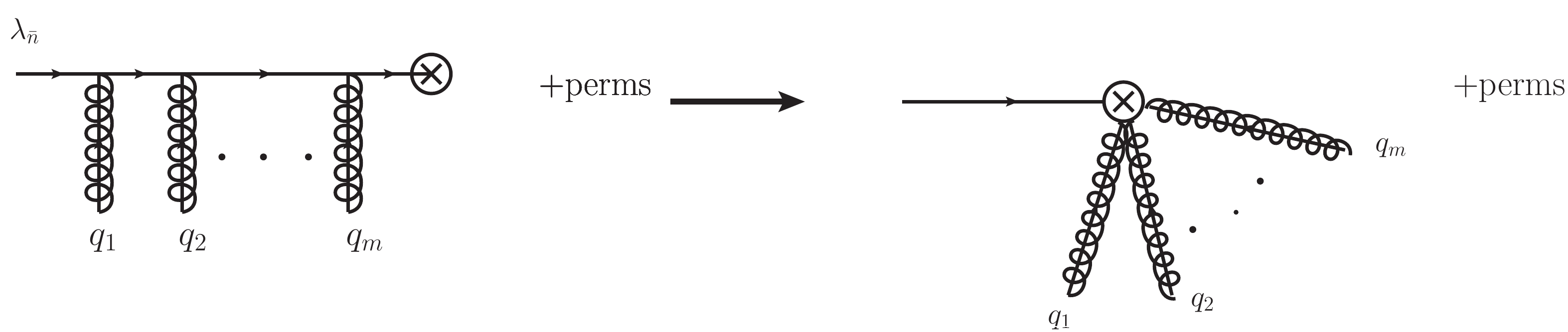}
\end{center}\vspace{-.5cm}
\caption{\baselineskip 3.0ex 
Feynman diagram in which $\lambda_{\bar{n}}$ emits collinear gluons in the $n$ direction. 
The intermediate states are off shell to be integrated out,
and the result produces the collinear Wilson line $\W_n$.}
\label{cowilson}
\end{figure}

Under the collinear gauge transformation $\mathcal{U}_c$, 
$\lambda_n$ transforms as $\mathcal{U}_c \lambda_n$, while $\W_n$ transforms as $\mathcal{U}_c \W_n$. Therefore the combination 
$\chi_n = \W_n^{\dagger} \lambda_n$ is invariant under collinear gauge transformation. 
This is one of the basic building blocks in constructing collinear gauge invariant operators.
The construction of the collinear Wilson line can be viewed as follows: 
If a particle not in the $n$ direction, say, a fermion in the $\overline{n}$ direction,
emits $n$-collinear gauge particles, the intermediate fermion states are off-shell and 
they should be integrated out. This
situation is illustrated in Fig.~\ref{cowilson}. In fact, it does not matter whether the particle which emits 
$n$-collinear gauge fields is
a collinear fermion in the $\overline{n}$ direction, or any other particle. 
What matters is to consider a particle not in the $n$ direction,
which emits $n$-collinear gauge particles and the off-shell intermediate states generate the collinear Wilson line. 

Integrating out the off-shell intermediate states,
the Feynman diagram in Fig.~\ref{cowilson} reads
\begin{equation} \label{wnexp}
\W_n = \sum_{m=0} \sum_{\mathrm{perm}} \frac{(-g)^m}{m!} \frac{\overline{n}\cdot A_{n,q_1}^{a_1}
\cdots \overline{n} \cdot A_{n,q_m}^{a_m}}{\overline{n}\cdot q_1 \overline{n}\cdot (q_1 +q_2) \cdots \overline{n}
\cdot \Bigl( \sum_{i=1}^m  q_i \Bigr)} T^{a_m} \cdots T^{a_1}.
\end{equation}
Since the generators $T_a$ are in the adjoint representation, the matrix element of $\W_n$ is given by 
\begin{eqnarray} \label{cowils}
(\W_n)_{ab} &=& \delta_{ab} \\
&+&\sum_{m=1} \sum_{\mathrm{perm}} \frac{(ig)^m}{m!} \frac{\overline{n}\cdot A_{n,q_1}^{a_1}
\cdots \overline{n} \cdot A_{n,q_m}^{a_m}}{\overline{n}\cdot q_1 \overline{n}\cdot (q_1 +q_2) \cdots \overline{n}
\cdot \Bigl( \sum_{i=1}^m  q_i \Bigr)} f^{a_m a x_{m-1}} \cdots f^{a_2x_2 x_1}f^{a_1 x_1 b}. \nonumber
\end{eqnarray}
Eq.~(\ref{wnexp}) is the explicit expansion of Eq.~(\ref{colwil}). For collinear gluons and scalar particles,
since they are also in the adjoint representation, the expression for the collinear Wilson line $\W_n$ is the same. 
In coordinate space, $\W_n$ is related to the
Fourier transform of the path-ordered exponential
\begin{equation}
\W_n (x) = \mathrm{P} \exp \Bigl( ig\int_{-\infty}^x ds \overline{n}\cdot A_n^a (ns) T^a\Bigr). 
\end{equation}

With the use of the collinear Wilson line, the collinear Lagrangian for fermions can be made  
manifestly collinear gauge invariant.  First, note that the collinear Wilson line $\W_n$ in Eq.~(\ref{colwil}) satisfies 
the equation of motion
\begin{equation}
\Bigl[ (\overline{\mathcal{P}} + g\overline{n} \cdot \mathcal{A}_n)\W_n\Bigr] =0,
\end{equation}
where $\overline{\mathcal{P}} =\overline{n}\cdot \mathcal{P}$ and  the bracket means the operator acts
only inside the bracket. Using this, the following relation
\begin{equation}
f(\overline{\mathcal{P}} +g\overline{n}\cdot \mathcal{A}_n) =\W_n f(\overline{\mathcal{P}}) \W_n^{\dagger}
\end{equation}
holds for an arbitrary function $f(\overline{\mathcal{P}})$. 
Then the collinear Lagrangian at leading order in $\eta$ is written as
\begin{equation} \label{lagcom}
\mathcal{L}_{\lambda}^{(0)} = \frac{1}{2} \overline{\lambda}_n \frac{\overline{n}\cdot \overline{\sigma}}{2}
\Bigl(n\cdot i\mathcal{D}
+\sigma\cdot \mathcal{D}_{c\perp} \W_n
\frac{1}{\overline{\mathcal{P}}} \W_n^{\dagger} \overline{\sigma} \cdot 
\mathcal{D}_{c\perp}\Bigr) \lambda_n.
\end{equation}
The Lagrangian is manifestly invariant under the collinear gauge transformation 
$\lambda \rightarrow \mathcal{U}_c \lambda$, $\W_n \rightarrow \mathcal{U}_c \W_n$,

So far, the Lagrangian for collinear fermions and the Wilson line are described in the adjoint representation. 
We can express these quantities in the fundamental representation. 
The first term of the Lagrangian in Eq.~(\ref{lagcom}) can be separately written as
\begin{eqnarray}
&&\frac{1}{2} \overline{\lambda}_n^a \frac{\overline{n}\cdot \overline{\sigma}}{2} n\cdot i\partial \lambda_n^a
= \mathrm{Tr}\, \Bigl( \overline{\lambda}_n \frac{\overline{n}\cdot 
\overline{\sigma}}{2} n\cdot i\partial \lambda_n\Bigr),  \nonumber \\
&&\frac{1}{2} igf^{abc}\overline{\lambda}_n^a n\cdot A_n^b \lambda_n^c = 
g\overline{\lambda}_n^a n\cdot A_n^b \lambda_n^c \mathrm{Tr}\, t^a [t^b,t^c] = 
g\mathrm{Tr}\, \Bigl(\overline{\lambda}_n [n\cdot A_n , \lambda_n]\Bigr),
\end{eqnarray}
where $A_n^{\mu} = A_n^{\mu a}t^a$ and $\lambda_n = \lambda_n^a t^a$. Combining these terms,
the first term in the Lagrangian can be written as
\begin{equation}
\mathrm{Tr}\, \Bigl(\overline{\lambda}_n \frac{\overline{n}\cdot \overline{\sigma}}{2}
 n\cdot iD \lambda_n\Bigr).
\end{equation}

To convert the second term in Eq.~(\ref{lagcom}), let us introduce the collinear Wilson line defined as
\begin{equation}
W_n = \sum_{\mathrm{perm.}}\exp \Bigl[-g\frac{1}{\overline{n}\cdot \mathcal{P}} \overline{n}\cdot A_n^a t^a
\Bigr].
\end{equation}
Compared to $\W_n$,  the only difference is that the generators for the gauge fields are in the fundamental
representation $t^a$ in $W_n$. It is also related to the Fourier transform of the collinear Wilson line
 \begin{equation}
W_n (x) = \mathrm{P} \exp \Bigl( ig\int_{-\infty}^x ds \overline{n}\cdot A_n^a (ns) t^a\Bigr). 
\end{equation}
The adjoint representation can be defined in terms of the fundamental representation by
\begin{equation} \label{adfun}
W_n t^a W_n^{\dagger} = \W_n^{ba} t^b.
\end{equation}
Therefore the expression of the form $\W_n^{\dagger} f_n$, where $f_n$ is any field in the adjoint representation, 
can be written in terms of the fundamental representation as
\begin{equation}
\W_n^{\dagger} f_n = \W_n^{\dagger}t^b f_n^b = (\W_n^{\dagger})^{ab} t^a f_n^b= W_n^{\dagger} t^b f_n^b W_n 
=W_n^{\dagger} f_n W_n. 
\end{equation}

Now consider the block 
$\W_n^{\dagger} \overline{\sigma} \cdot \mathcal{D}_{c\perp} \lambda_n$, which can be written explicitly as
\begin{equation}
\Bigl(\W_n^{\dagger} \overline{\sigma} \cdot \mathcal{D}_{c\perp} \lambda_n \Bigr)^a
=(\W_n^{\dagger})^{ab} \overline{\sigma}\cdot \mathcal{P}_{\perp} \lambda_n ^b 
+(\W_n^{\dagger})^{ab} igf^{bcd} \overline{\sigma} \cdot A_{n\perp}^c \lambda_n^d.
\end{equation}
Multiplying $t^a$ on both sides, we obtain
\begin{eqnarray}
&&W_n^{\dagger} \overline{\sigma}\cdot \mathcal{P}_{\perp} \lambda_n W_n + W_n^{\dagger}
[t^c,t^d] W_n  g\overline{\sigma}\cdot  A_{n\perp}^c \lambda_n^d =
 W_n^{\dagger} \overline{\sigma}\cdot \mathcal{P}_{\perp} 
\lambda_n W_n + W_n^{\dagger} [g\overline{\sigma} \cdot A_{n\perp}, \lambda_n]W_n \nonumber \\
&&= W_n^{\dagger} \overline{\sigma}\cdot D_{c\perp}\lambda_n W_n.
\end{eqnarray}
Finally the Lagrangian in Eq.~(\ref{lagcom}) can be written in terms of the fundamental representation as
\begin{equation} \label{lagfun}
\mathcal{L}_{\lambda}^{(0)} = \mathrm{Tr}\, \Biggl[ W_n^{\dagger} \Biggl( \overline{\lambda}_n 
\frac{\overline{n}\cdot \overline{\sigma}}{2} \Bigl( n\cdot iD +\sigma\cdot D_{c\perp} W_n 
\frac{1}{\overline{\mathcal{P}}} W_n^{\dagger} \overline{\sigma}\cdot D_{c\perp}\Bigr) \lambda_n\Biggr) W_n\Biggr].
\end{equation}
The collinear Wilson lines outside the parenthesis cancel  due to the trace. For comparison, the corresponding Lagrangian
for QCD is given by
\begin{equation}
\mathcal{L}^{(0)}_{\mathrm{QCD}} = \overline{\xi}_n \frac{\FMslash{\overline{n}}}{2} 
\Bigl( n\cdot iD +\FMSlash{D}_{c\perp} W_n \frac{1}{\overline{\mathcal{P}}} W_n^{\dagger}
 \FMSlash{D}_{c\perp} \Bigr) \xi_n.
\end{equation}

The fermion-scalar interaction beginning at order $\eta$ can be expressed in terms of the collinear Wilson lines. 
According to the relation in Eq.~(\ref{nofs}), the SCET Lagrangian from the full-theory interaction is  given by
\begin{equation}
\mathcal{L}_{\mathrm{fs}} =\left\{ \begin{array}{l}
 -ig \mathrm{Tr} \, \Bigl( \lambda_i [\lambda_j,\phi^{ij}]\Bigr) \longrightarrow ig \mathrm{Tr} \, 
\Bigl( \lambda_{j,n} [\lambda_{i,\bar{n}}, \phi^{ij}]\Bigr) \\
\displaystyle \frac{g}{2} f^{abc} \lambda_i^a \lambda_j^b \phi^{ij,c}\longrightarrow
\frac{g}{2} f^{abc} \lambda_{j,n}^b \lambda_{i,\bar{n}}^a \phi^{ij,c},
\end{array}
\right.
\end{equation}
in the fundamental and the adjoint representations respectively. Here the scalar field can be either 
$n$-collinear or usoft.

In the adjoint representation the small component $\lambda_{i,\bar{n}}$ at leading order from Eq.~(\ref{subad})
is given by
\begin{equation} \label{smallad}
\lambda_{i,\bar{n}}^a = -\frac{\overline{n}\cdot \sigma}{2} \Bigl(\W_n \frac{1}{\overline{\mathcal{P}}} 
\W_n^{\dagger} \overline{\sigma} \cdot \mathcal{D}_{c\perp} \lambda_{i,n}\Bigr)^a,
\end{equation}
and the interaction is written as
\begin{equation}
\mathcal{L}_{\mathrm{fs}} = -\frac{g}{2} f^{abc} \lambda_{j,n}^b
\frac{\overline{n}\cdot \sigma}{2} \Bigl(\W_n \frac{1}{\overline{\mathcal{P}}} 
\W_n^{\dagger} \overline{\sigma} \cdot \mathcal{D}_{c\perp} \lambda_{i,n}\Bigr)^a \phi^{ij,c}.
\end{equation}
The small component $\lambda_{\bar{n}}$ in the fundamental representation can be written, using Eqs.~(\ref{smallad})
and (\ref{adfun}), as
\begin{equation}
W_n^{\dagger} \lambda_{\bar{n}} W_n =  -\frac{\overline{n}\cdot \sigma}{2} \frac{1}{\overline{\mathcal{P}}} 
W_n^{\dagger} \overline{\sigma} \cdot D_{c\perp} \lambda_n W_n.
\end{equation}
And the interaction in the fundamental representation is written as
\begin{eqnarray}
\mathcal{L}_{\mathrm{fs}} &=& ig \mathrm{Tr} \, 
\Bigl( W_n^{\dagger} \lambda_{j,n} W_n [W_n^{\dagger}\lambda_{i,\bar{n}} W_n, W_n^{\dagger} 
\phi^{ij} W_n]\Bigr)  \nonumber \\
&=&-ig \mathrm{Tr} \, \Biggl( W_n^{\dagger} \lambda_{j,n} W_n \Biggl[ \Bigl[
\frac{1}{\overline{\mathcal{P}}} \frac{\overline{n}\cdot \sigma}{2} 
W_n^{\dagger} \overline{\sigma} \cdot D_{c\perp} \lambda_{i,n} W_n\Bigr], W_n^{\dagger} 
\phi^{ij} W_n\Biggr] \Biggr).
\end{eqnarray}

The scalar-fermion interaction involves a small component $\lambda_{\bar{n}}$. But according to the power counting,
this interaction is of the same order as the other leading Lagrangian. Collinear particles scale as $\eta$, usoft 
particles scale as $\eta^2$, and collinear gauge particles scale as in Eq.~(\ref{gauge}). This scaling behavior is obtained
by considering $\int d^4 x \mathcal{L}$, where the volume element $d^4 x$ scales as $\eta^{-4}$ ($\eta^{-8}$)
for collinear (usoft) particles, and $\mathcal{L}$ scales as $\eta^4$ ($\eta^8$) in each case. The effective
Lagrangian for collinear fermions in Eq.~(\ref{lagfun}) scales as $\eta^4$, and $\mathcal{L}_{\mathrm{fs}}$ 
is also of order $\eta^4$. Therefore  $\mathcal{L}_{\mathrm{fs}}$ is also a leading Lagrangian in SCET.

\section{Usoft factorization\label{ufac}}      
\begin{figure}[t]
\begin{center} 
\includegraphics[height=3.cm]{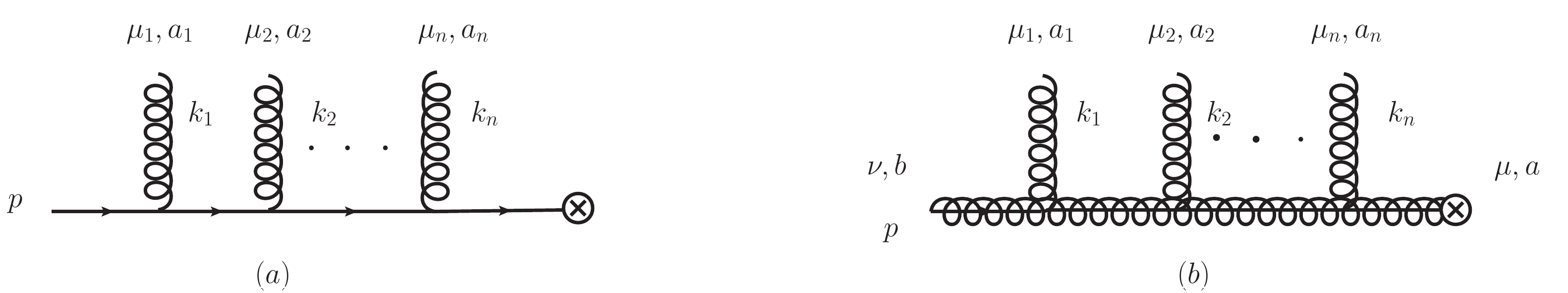}
\end{center}\vspace{-.5cm}
\caption{\baselineskip 3.0ex 
Feynman diagrams in which usoft gauge particles are attached to (a) a collinear fermion  and (b) a collinear gauge particle. 
The color factors both for the fermions, the scalars and the gauge particles are the same. The diagram with a collinear 
scalar particle is omitted. }
\label{usoft}
\end{figure}

One of the most interesting features in SCET is that SCET is formulated such that collinear particles are decoupled from usoft
interactions. This can be achieved by redefining the collinear fields in terms of the usoft Wilson lines.
Consider the interaction of the collinear fields with usoft background gauge fields, and the relevant
Feynman diagrams are shown in Fig.~\ref{usoft}. The sum of the diagrams which couple usoft gauge particles to the collinear 
fields is given as
\begin{equation} \label{ufac1}
f_n^a = \Y_n^{ab} f_n^{(0) b},
\end{equation}
where $f_n = \lambda_n,  A_n^{\mu}, \phi_n$ denote collinear fields, and $\Y^{ab}$ is written as
\begin{equation}
\Y_n^{ab} =\delta^{ab} +\sum_{m=1}^{\infty} \sum_{\mathrm{perm}} \frac{(ig)^m}{m!}
\frac{n\cdot A_{\mathrm{us}}^{a_1} \cdots n\cdot A_{\mathrm{us}}^{a_m}}{n\cdot k_1 n\cdot (k_1 +k_2)
\cdots n\cdot \Bigl(\sum_{i=1}^m k_i\Bigr)}f^{a_m a x_{m-1}} \cdots f^{a_2x_2 x_1}f^{a_1 x_1 b}.
\end{equation}
It is also the explicit expansion of the usoft Wilson line
\begin{equation}
\Y_n =\sum_{\mathrm{perm.}} \exp \Bigl[-g \frac{1}{n\cdot \mathcal{R}} n\cdot \mathcal{A}_{\mathrm{us}} \Bigr], 
\end{equation}
where $\mathcal{A}_{\mathrm{us}} = A_{\mathrm{us}}^a T^a$, and $n\cdot \mathcal{R}$ is the operator extracting the
momentum $n\cdot p$.
Note that $\Y_n$ has a similar structure compared to $\W_n$ as far as the color factors are concerned because
all the particles are in the adjoint representation, but
the projection of the gauge fields is in the $n$ direction, not $\overline{n}$ in contrast to the case of $\W_n$.
$\Y_n$ is related to the Fourier transform of the usoft Wilson line in the adjoint representation
\begin{equation}
\Y_n^{ab} (x) =\Biggl[ \mathrm{P} \exp\Bigl( ig \int_{-\infty}^x ds n\cdot A_{\mathrm{us}}^c (ns) T^c\Bigr) 
\Biggr]^{ab}.
\end{equation}

It is useful to introduce the usoft Wilson line in the fundamental representation, which is defined as 
\begin{equation}
Y_n =1 +\sum_{m=1}^{\infty} \sum_{\mathrm{perm}} \frac{(-g)^m}{m!}
\frac{n\cdot A_{\mathrm{us}}^{a_1} \cdots n\cdot A_{\mathrm{us}}^{a_m}}{n\cdot k_1 n\cdot (k_1 +k_2)
\cdots n\cdot \Bigl(\sum_{i=1}^m k_i\Bigr)} t^{a_m} \cdots t^{a_1},
\end{equation}
which is obtained from $\Y_n$ by replacing the adjoint representation of the $SU(N)$ color generators
$(T^a)^{bc}=-if^{abc}$ by the fundamental representation $t^a$. It is related to the Fourier transform of 
the usoft Wilson line
\begin{equation}
Y_n (x) = \mathrm{P} \exp\Bigl( ig \int_{-\infty}^x ds n\cdot A_{\mathrm{us}}^c (ns) t^c\Bigr).
\end{equation}

The relation between the adjoint and the fundamental representations can be given by
\begin{equation}
Y_n t^a Y_n^{\dagger} =\Y_n^{ba} t^b,
\end{equation}
from which we obtain that
\begin{equation} \label{ufac2}
f_n = f_n^b t^b = f_n^{(0)a} \Y_n^{ba} t^b = f_n^{(0)a} Y_n t^a Y_n^{\dagger} = Y_n f_n^{(0)} Y_n^{\dagger}. 
\end{equation}
This is similar to the case involving $\W_n$ because, again, all the particles are in the adjoint representation.
Physically $f_n$ is regarded as the collinear field immersed in the cloud of usoft gauge fields, and after factoring out the usoft
contribution using the usoft Wilson line, $f_n^{(0)}$ is the collinear field decoupled from the usoft interaction. 

We can express the collinear Lagrangian in terms of the redefined collinear fermion fields $\lambda_n^{(0)}$
and $A_n^{(0)\mu}$. 
Since $A_n^{\mu} = YA_n^{(0)\mu} Y^{\dagger}$, it follows that
\begin{equation}
W_n =\Bigl[\sum_{\mathrm{perm}} \exp \Bigl( -g\frac{1}{\overline{\mathcal{P}}} Y\overline{n}\cdot
A_n^{(0)} Y^{\dagger}\Bigr) \Bigr] = YW_n^{(0)}Y^{\dagger},
\end{equation}
which also shows how usoft gauge particles couple to the collinear Wilson line. Starting with the collinear fermion 
Lagrangian in Eq.~(\ref{lagfun}), we obtain
\begin{eqnarray}
\mathcal{L}_{\lambda}^{(0)} &=& \mathrm{Tr}\, \Biggl(  \overline{\lambda}_n^{(0)}
\frac{\overline{n}\cdot \overline{\sigma}}{2} \Bigl[ n\cdot iD_c^{(0)} +\sigma\cdot D_{c\perp}^{(0)} W_n^{(0)} 
\frac{1}{\overline{\mathcal{P}}}W_n^{(0)\dagger} \overline{\sigma}\cdot D_{c\perp}^{(0)} 
\Bigr] \lambda_n^{(0)}\Biggr),
\end{eqnarray}
where we use the facts that $\mathcal{P}_{\perp}^{\mu}$ commutes with $Y$ and $Y^{\dagger}n\cdot D_{\mathrm{us}}
Y = n\cdot \partial$ since $n\cdot D_{\mathrm{us}}Y=0$. This is the final collinear Lagrangian in which the collinear 
fermion is decoupled from the usoft interaction. Similarly, the collinear scalar Lagrangian is given by
\begin{equation}
\mathcal{L}_s = \mathrm{Tr}\, \Biggl(\phi_n^{(0)*} \Bigl( n\cdot iD_c+\frac{\mathcal{P}_{\perp}^2}{\overline{n}\cdot
\mathcal{P}} \Bigr)\phi_n^{(0)} \Biggr).
\end{equation}
From now on, we drop the superscript ${}^{(0)}$ and we have established the collinear Lagrangian at leading order, 
which is decoupled from the usoft interaction.

Note that we do not have to include the effect of the scalar particle emissions for the usoft factorization of
the SCET Lagrangian at leading order. 
If a scalar particle
interacts with collinear fermions at leading order, we may have to include the emissions of collinear or usoft scalar 
particles from collinear fermions 
to all orders in $g$ to extract the quantities similar to collinear or usoft Wilson lines. 
However, if there are $m$ scalar particles emitted from a collinear fermion, it has the dependence of $(\phi_n)^m$ or 
$(\phi_{\mathrm{us}})^m$ for collinear and usoft scalar particles respectively and they are suppressed by $\eta^m$
or $\eta^{2m}$. Therefore the scalar particle does not interact with collinear fermions at leading order whether 
the scalar is collinear or usoft. Physically this is related to the chirality flip due to the scalar interaction. If a scalar particle 
is emitted from a collinear fermion, the fermion becomes an antifermion and the chirality is flipped. Chirality flip can occur
only for massive particles, hence it does not occur for massless fermions. The emission of collinear or usoft scalar particles
becomes more subleading as the number of the emitted scalar particles increases, and we can safely discard them at 
leading order. This makes the structure of the effective theory
simple.

\section{Application\label{appl}}
There may be various applications, and we mention two possible applications here 
in applying the SCET formulation of the $\mathcal{N}=4$ SYM
theory. First we can construct gauge-invariant operators in SCET, and factorize the collinear and the usoft parts. 
Each part in turn can be computed
using perturbation theory. Secondly, we can consider scattering amplitudes such as $gg\rightarrow gg$, 
$gg \rightarrow \overline{\lambda} \lambda$, or $\lambda \lambda \rightarrow \lambda \lambda$ and study the
divergence structure of the scattering amplitudes. The factorization of the collinear part and the usoft part is 
critical and interesting, since we can keep track of the origins of the divergences in calculating each part.  
Here we illustrate an example to present
the basic ideas about how to apply the techniques of SCET for the radiative corrections of a current operator,
and leave the study of scattering amplitudes in a forthcoming paper.

Consider an operator in the full theory, e. g., the back-to-back collinear fermion current operator of the form 
$J^{\mu} =\overline{\lambda} \overline{\sigma}^{\mu} \lambda$. In SCET, 
the corresponding operator is given by
\begin{equation}
J_c^{\mu}=C(Q,\mu) \overline{\lambda}_{\bar{n}} \W_{\bar{n}} \overline{\sigma}_{\perp}^{\mu} 
\W_n^{\dagger} \lambda_n= 2C(Q,\mu) 
\mathrm{Tr}\, \Bigl( W_{\bar{n}}^{\dagger}\overline{\lambda}_{\bar{n}} W_{\bar{n}} 
\overline{\sigma}_{\perp}^{\mu} W_n^{\dagger} \lambda_n W_n\Bigr),
\end{equation}
where $C(Q,\mu)$ is the Wilson coefficient by matching the full theory onto the SCET at some large
scale $Q$. The current operator is collinear gauge invariant by attaching the collinear Wilson lines. The usoft
interaction can be obtained after redefining the collinear fields, but here we will use the current operator without
the usoft Wilson line, and consider the usoft interaction employing the Feynman rules given in Fig.~\ref{feynman}. 

\begin{figure}[b]
\begin{center} 
\includegraphics[height=2.cm]{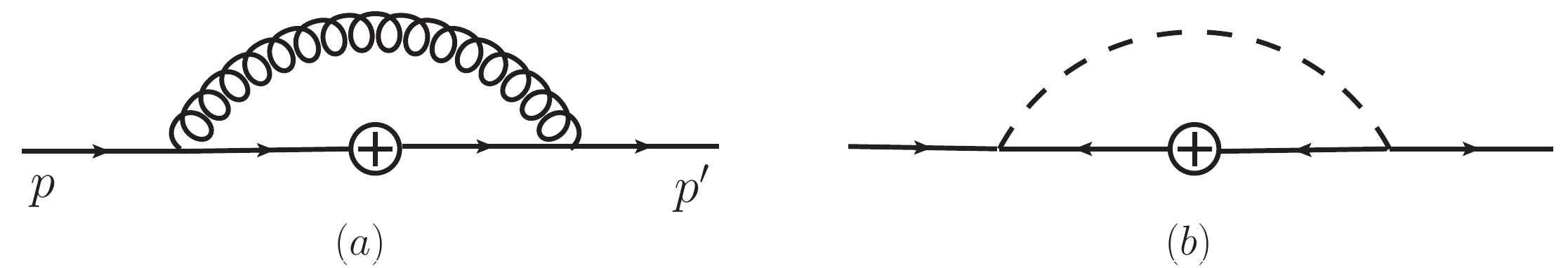}
\end{center}\vspace{-.5cm}
\caption{\baselineskip 3.0ex 
Feynman diagrams for the vertex corrections at one loop in the full theory with the exchange of 
(a) a gauge particle, and (b) a scalar particle.}
\label{fullcurrent}
\end{figure}

In order to see the consistency of the effective theory, we compute the infrared divergent part of the full theory in the vertex 
correction and the finite part constitutes the Wilson coefficient. We use the dimensional regularization for 
the UV divergence with the spacetime dimension $D=4-2\epsilon$, and take the nonzero offshellness of the external particles
as infrared cutoffs. 
The Feynman diagrams for the vertex corrections in the full theory are shown in Fig.~\ref{fullcurrent}. It turns out that
Fig.~\ref{fullcurrent} (a) gives the same vertex correction as in QCD except the color factor from $C_F = (N^2 -1)/(2N)$
to $C_A=N$ where $N$ is the number of colors. Fig.~\ref{fullcurrent} (b) is the new contribution, which is absent in QCD. 
In both diagrams there are ultraviolet divergences. But they are cancelled by the wave function renormalization, which is not 
shown in Fig.~\ref{fullcurrent}. This is due to the current conservation. Fig.~\ref{fullcurrent} (a) has IR divergences, but 
Fig.~\ref{fullcurrent} (b) is infrared finite. The IR divergence in the full theory should be reproduced in SCET, which we will
explicitly show below. 

The explicit computation of Fig.~\ref{fullcurrent} is given as
\begin{equation} \label{fulver}
M_{\mathrm{full}} =-\frac{g^2 C_A}{16\pi^2} \overline{\sigma}^{\mu} 
\Bigl[ 2 \ln \frac{-p^2}{Q^2} \ln \frac{-p'^2}{Q^2}
+2 \ln \frac{-p^2}{Q^2} +2\ln \frac{-p'^2}{Q^2} -3+\frac{2\pi^2}{3} +\ln\frac{Q^2}{\mu^2}\Bigr],
\end{equation}
where $q^2= (p-p')^2 =-2p\cdot p' = -\overline{n}\cdot p n\cdot p' =-Q^2$. The first term is the IR divergence at one loop
in the full theory. The second and the third terms contain $\ln p^2$ IR divergences, but they are cancelled by analogous 
IR divergence in bremsstrahlung process in computing the scattering cross section.

\begin{figure}[t]
\begin{center} 
\includegraphics[height=2.cm]{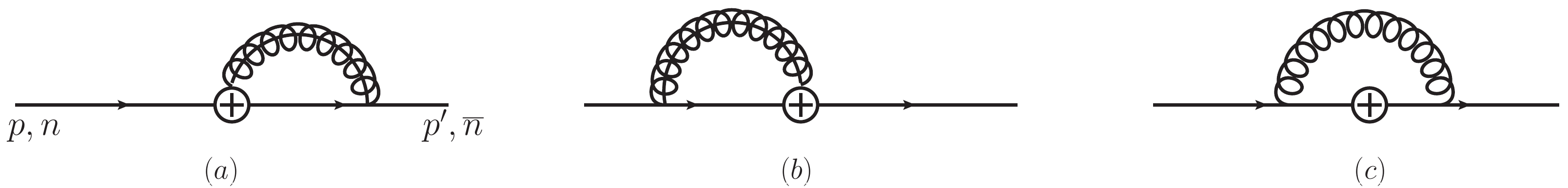}
\end{center}\vspace{-.5cm}
\caption{\baselineskip 3.0ex 
Feynman diagrams for the vertex corrections at one loop in SCET with the exchange of (a) a 
$\overline{n}$-collinear gauge particle, (b) a $n$-collinear gauge particle and (c) an usoft gauge particle.}
\label{scetcurrent}
\end{figure}

In SCET, there is no fermion-scalar interaction at leading order, 
and there are only collinear and usoft gauge interactions. 
The Feynman diagrams for the vertex correction in SCET at one loop are shown
in Fig.~\ref{scetcurrent}. 
The collinear contributions from Fig.~\ref{scetcurrent} (a) and (b) are given as
\begin{equation}
M_c = -\frac{g^2 C_A}{16\pi^2}\overline{\sigma}^{\mu} 
\Bigl[ -\frac{4}{\euv^2} -\frac{2}{\euv} \Bigl(\ln \frac{\mu^2}{-p^2}
+\ln \frac{\mu^2}{-p'^2}\Bigr) -\ln^2 \frac{\mu^2}{-p^2} -\ln^2 \frac{\mu^2}{-p'^2} \Bigr]. 
\end{equation}
The usoft contribution from Fig.~\ref{scetcurrent} (c) is given by 
\begin{equation}
M_{\mathrm{us}} = -\frac{g^2 C_A}{16\pi^2}\overline{\sigma}^{\mu}  
\Bigl[ \frac{2}{\euv^2} +\frac{2}{\euv} 
\Bigl(\ln \frac{\mu n\cdot p'}{-p'^2} +\ln \frac{\mu \overline{n}\cdot p}{-p^2}\Bigr) 
+\ln^2 \frac{(-p^2)(-p'^2)}{\mu \overline{n}\cdot p \mu n\cdot p'}\Bigr].
\end{equation}
The detailed computation is presented in Appendix.
Note that each of the collinear contribution and the usoft contribution contains UV divergences, and there is also a troublesome
quantity $(1/\euv)\ln (-p^2/\mu^2)$ in each contribution, which is a mixture of UV and IR divergences. However,
the sum of the two contributions is free of this term and the UV and the IR divergences are separated. 
The overall contribution is written as
\begin{equation} \label{scetres}
M=M_c +M_{\mathrm{us}} = -\frac{g^2 C_A}{16\pi^2} \overline{\sigma}^{\mu} 
\Bigl[ -\frac{2}{\euv^2} -\frac{2}{\euv} \ln \frac{\mu^2}{Q^2}
-\ln^2 \frac{\mu^2}{Q^2} +2\ln \frac{-p^2}{Q^2} \ln\frac{-p'^2}{Q^2}\Bigr].
\end{equation}
The first two terms contribute to the anomalous dimension of the operator, and the third term contributes to the Wilson
coefficient along with the finite terms in $M_{\mathrm{full}}$. The last term is the IR divergence. 
Comparing with the full theory calculation
in Eq.~(\ref{fulver}), the IR divergence of the full theory is exactly reproduced in SCET.

\section{Conclusion and Outlook}
We have constructed the SCET for $\mathcal{N}=4$ SYM theory. This effective theory shows many 
interesting features. First of all, all the particles in this theory are in the adjoint representation of the $SU(N)$ gauge group,
simplifying the structure of the theory.  In order to describe the fermion sector in Weyl representation, we introduce
the appropriate projection operators to construct the Lagrangian. Using the gauge transformation properties of the fields,
the Lagrangian and any operators can be constructed in a
collinear and usoft gauge-invariant way.  These can be expressed either in the adjoint or in the fundamental representations. 
One striking feature is that there is no interaction between collinear fermions and
collinear/usoft scalar particles at leading order, but it begins with order $\eta$. 
Due to this fact, the redefinition of the collinear fields to decouple the usoft interaction
is accomplished only by the usoft Wilson lines from the usoft gauge fields. 

We have shown how to renormalize a current operator as an example, and it is easy to trace the origins of the divergences, 
whether they come from collinear or usoft parts. One-loop computations may be too simple to see if SCET 
serves better than the full theory in some respects,
and we have to consider radiative corrections at higher loops. Since the example deals with one-loop corrections, 
all the radiative corrections are proportional to 't Hooft coupling $\lambda = g^2 N$. It would be interesting to see if 
planar and nonplanar diagrams can be organized conveniently in SCET. 

This paper is the first step to consider $\mathcal{N}=4$ SYM in terms of SCET, and it opens many questions to be answered.
One intensive field of interest is high-energy scattering amplitudes. In the full $\mathcal{N}=4$ SYM 
theory, gluon scattering amplitudes at higher-loops and the divergence structure are actively investigated. And it will be
interesting to view from different perspectives to understand the behavior and the divergence structure
of high-energy scattering amplitudes. Another field is to consider anomalous dimensions of some operators. Of course, the 
results in the full theory are well beyond one loop and there is a large gap at the moment between 
the full theory \cite{Bern:2006ew} and SCET.
Also there exists duality between Wilson loops and gluon amplitudes \cite{Drummond:2007cf}. The leading IR divergences
of gluon amplitudes are equivalent to the leading UV divergences of Wilson loops. 
The collinear and usoft Wilson lines derived here can be a starting point to consider the duality relation in view of SCET.

In addition to applying the ideas of SCET to known results in the full theory to understand the structure better, 
SCET itself poses several interesting questions. For example, in the SCET for QCD,
the classification of the gauge transformations into collinear and usoft gauge transformations is useful in considering the
structure of operators, and SCET offers richer gauge symmetries than the original gauge symmetry. Supersymmetry is an additional
symmetry of the theory. Possibly the supersymmetry
transformations may be classified into different classes, under which particles transform in a nontrivial way. 
One of the supersymmetry algebra is given by $\{ Q_{\alpha}, Q_{\dot{\alpha}}\} 
= 2\sigma_{\alpha \dot{\alpha}}^{\mu} P_{\mu}$, which depends on the momentum operator.
If we find subclasses of supersymmetry transformations with respect to the momentum operator of definite power counting,
it might help understand the structure of the full theory better.  Combined with superconformal property of the theory, 
the SCET will show a diverse structure of the theory.

\begin{acknowledgments}
Both authors are  supported  by Mid-career Researcher Program through NRF grant funded by the MEST (2010-0027811). 
J. Y. Lee is supported in part  by Basic Science Research Program through the NRF of Korea funded by the MEST (2010-0012779).  
\end{acknowledgments}

\appendix

\section{Explicit calculation of vertex corrections at one loop in SCET}

\begin{figure}[b]
\begin{center} 
\includegraphics[height=2.5cm]{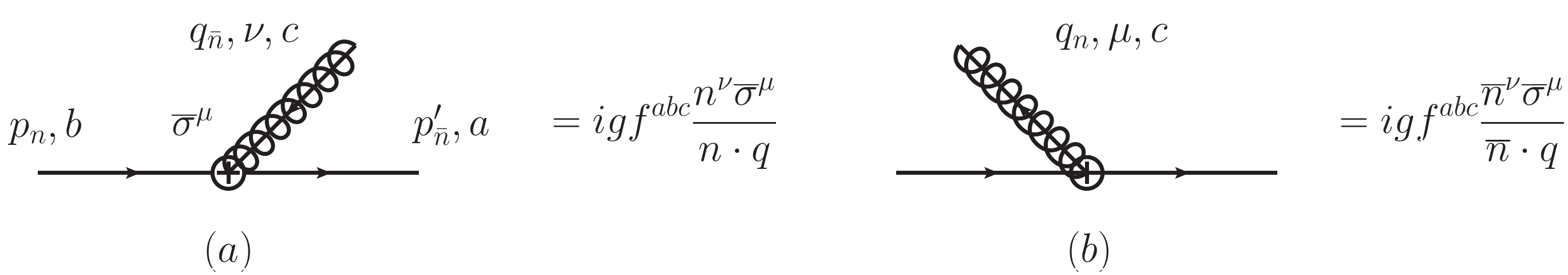}
\end{center}\vspace{-.5cm}
\caption{\baselineskip 3.0ex 
Feynman rules for the current from the collinear Wilson line. (a) $\overline{n}$-collinear gauge particle,
 (b) $n$-collinear gauge particle.}
\label{cfeyn}
\end{figure}

The Feynman rules for the vertex of the current from the collinear Wilson line are shown in Fig.~\ref{cfeyn}.
Fig.~\ref{scetcurrent} (a) is written as
\begin{equation}\label{naicol}
M_a = -2ig^2 C_A \overline{\sigma}^{\mu} \int \frac{d^D l}{(2\pi)^D} \frac{n\cdot (l+p')}{l^2 (l+p')^2 n\cdot l}, 
\end{equation}
where the $\sigma$ matrices other than $\overline{\sigma}^{\mu}$ becomes a projection operator in the $n$ direction. 
 It is given by
\begin{equation}
M_a = -\frac{g^2 C_A}{16\pi^2} \overline{\sigma}^{\mu} \Bigl[ -\frac{2}{\euv\eir} -\frac{2}{\euv}
\ln \frac{\mu}{n\cdot p'} +\frac{2}{\eir} \ln \frac{\mu}{n\cdot p'} -\frac{2}{\eir} \ln \frac{\mu^2}{-p'^2}
- \ln^2 \frac{\mu^2}{-p^2}\Bigr].
\end{equation}
In performing the loop integration, though $p'^2$ acts as an IR cutoff, there appear poles in $1/\eir$. 

One caveat is that
the collinear loop integral covers the usoft region, which should be avoided since it is taken care of by the usoft interaction. 
The corresponding contribution from the usoft limit in the collinear integral is removed 
by the zero-bin subtraction \cite{Manohar:2006nz}. 
It is obtained by modifying the scaling behavior of the loop momentum in Eq.~(\ref{naicol}). The collinear momenta scale as
$(n\cdot l, l_{\perp}, \overline{n} \cdot l) \sim (n\cdot p', p'_{\perp}, \overline{n} \cdot p')\sim Q(1,\eta, \eta^2)$,
but the zero-bin region corresponds to the momentum scaling $(n\cdot l, l_{\perp}, \overline{n} \cdot l) \sim Q(\eta^2,
\eta^2,\eta^2)$. With this power counting, the zero-bin contribution is written as
\begin{eqnarray}
M_a^{(0)} &=& -2ig^2 C_A \overline{\sigma}^{\mu} \int \frac{d^D l}{(2\pi)^D} \frac{n\cdot p'}{l^2 
(n\cdot p' \overline{n}\cdot l  +p'^2) n\cdot l} \nonumber \\
&=& -\frac{g^2 C_A}{16\pi^2} \overline{\sigma}^{\mu}\Bigl[ \frac{2}{\euv^2} 
-\frac{2}{\euv\eir} +\Bigl(\frac{2}{\euv} -\frac{2}{\eir} \Bigr)
\Bigl( \ln \frac{\mu^2}{-p'^2} -\ln \frac{\mu}{n\cdot p'}\Bigr)\Bigr],
\end{eqnarray}
where $p'^2$ enters as the IR cutoff. Finally the collinear contribution is given as
\begin{eqnarray}
\tilde{M}_a &=&M_a -M_a^{(0)} = -\frac{\alpha_s C_A}{4\pi} \overline{\sigma}^{\mu} 
\Bigl[-\frac{2}{\euv^2} -\frac{2}{\euv} \ln \frac{\mu^2}{-p'^2} -\ln^2 \frac{\mu^2}{-p'^2}\Bigr], \nonumber \\
\tilde{M}_b &=& -\frac{g^2 C_A}{16\pi^2} \overline{\sigma}^{\mu} 
\Bigl[-\frac{2}{\euv^2} -\frac{2}{\euv} \ln \frac{\mu^2}{-p^2} -\ln^2 \frac{\mu^2}{-p^2}\Bigr],
\end{eqnarray}
where we can proceed in the same way for the $n$-collinear loop integral appearing $\tilde{M}_b$ 
in Fig.~\ref{scetcurrent} (b) 
with the replacement of $p'^2$ by $p^2$ in the result.

The usoft contribution in Fig.~\ref{scetcurrent} (c) is given by
\begin{eqnarray}
M_c &=& -2ig^2 C_A \overline{\sigma}^{\mu} \int \frac{d^D l}{(2\pi)^D} 
\frac{1}{l^2 (\overline{n}\cdot l +p'^2/n\cdot p') (n\cdot l +p^2/\overline{n}\cdot p)} \nonumber \\
&=& -\frac{g^2 C_A}{16\pi^2} \overline{\sigma}^{\mu} \Bigl[\frac{2}{\euv^2} -\frac{2}{\euv}
\Bigl(\ln \frac{n\cdot p'}{\mu} +\ln \frac{\overline{n}\cdot p}{\mu}\Bigr) 
+\ln^2 \frac{(-p'^2)(-p^2)}{\mu^2 n\cdot p'\overline{n}\cdot p}\Bigr],
\end{eqnarray}
where $p^2$ and $p'^2$ act as IR regulators. The overall contribution in SCET is given as
\begin{equation}
M_{\mathrm{SCET}} = \tilde{M}_a +\tilde{M}_b +M_c = -\frac{g^2 C_A}{16\pi^2} \overline{\sigma}^{\mu} 
\Bigl[-\frac{2}{\euv^2} -\frac{2}{\euv} \ln \frac{\mu^2}{Q^2}-\ln^2 \frac{\mu^2}{Q^2} +2\ln \frac{-p^2}{Q^2} 
\ln \frac{-p'^2}{Q^2}\Bigr],
\end{equation}
which is Eq.~(\ref{scetres}).

\end{document}